\documentstyle[12pt]{article}
\input psfig
\def\fn{ \baselineskip = 0pt
\vbox{\hbox{\hspace*{3pt}\tiny $\circ$}\hbox{$f$}} 
\baselineskip = 12pt\!}

\def\R{\mathrm{ I\kern-.1567em R}}
\def\be{\begin{equation}}
\def\ee{\end{equation}}
\def\bea{\begin{eqnarray}}
\def\eea{\end{eqnarray}}
\def\m{\mu}
\def\l{\lambda}
\def\r{\rho}
\def\vo{\sqrt{1+|v|^2}}

\begin{document}

\title{Critical collapse of collisionless matter---a 
numerical investigation}
\author{Gerhard Rein\\
        Mathematisches Institut der Universit\"at M\"unchen\\
        Theresienstr.\ 39, 80333 M\"unchen, Germany\\
	Alan D.~Rendall\\
	Max-Planck-Institut f\"ur Gravitationsphysik\\
	Schlaatzweg 1, 14473 Potsdam, Germany\\
	Jack Schaeffer\\
	Department of Mathematics\\
	Carnegie-Mellon University, Pittsburgh, PA 15213, USA}
\date{}
\maketitle
\begin{abstract}
In recent years the threshold of black hole formation in spherically symmetric
gravitational collapse has been studied for a variety of matter models. In 
this paper the corresponding issue is investigated for a matter model 
significantly different from those considered so far in this context. We study
the transition from dispersion to black hole formation in the collapse of
collisionless matter when the initial data is scaled. This is done by means
of a numerical code similar to those commonly used in plasma physics. The
result is that for the initial data for which the solutions were computed,
most of the matter falls into the black hole whenever a black hole is
formed. This results in a discontinuity in the mass of the black hole at
the onset of black hole formation. 
\end{abstract}

\section{Introduction}
\setcounter{equation}{0}
The gravitational collapse of a localized concentration of matter to a
black hole is a central theme in general relativity. 
Even in the simplest case of spherically symmetric collapse 
much remains to be learned. When no matter is present, i.e. 
in the case of the vacuum Einstein equations, there
is no collapse in spherical symmetry, because of Birkhoff's 
theorem. Thus it is necessary, in order to obtain information 
about gravitational collapse by studying the spherically symmetric 
case, to choose a matter model. A simple choice, which has 
provided valuable insights, is that of a massless
(real, minimally coupled) scalar field. A deep mathematical 
investigation of spherically symmetric solutions of the 
Einstein-scalar field system has been carried out by Christodoulou. 
Some of his results will now be discussed.

In \cite{chr86} it was shown that for sufficiently small initial
data the fields disperse to infinity at large times. A result for 
the case of large data was proved in \cite{chr87}. For any solution 
it is possible to define a number $M$ with the property that in the 
region $r>2M$ (where $r$ is the area radius) the solution approaches 
the Schwarzschild solution of mass $M$ at large times. The physical 
interpretation is that the system has collapsed to form a black hole 
of mass $M$. Of course $M=0$ in the case that the field disperses. 
The dichotomy between dispersion and black hole formation still 
leaves the question unanswered, which data result in
which of these two outcomes, except for small data. In \cite{chr91}
Christodoulou gave a sufficient condition on initial data to ensure 
that $M>0$ for the corresponding solution, although this criterion 
is not very practical.

For large data the above results only describe the structure of the 
solution at sufficiently large radius and the internal structure is 
left open. In \cite{chr94} it was proved that there exist data 
leading to the formation of a naked singularity. The negative 
implications of this for the cosmic censorship hypothesis are limited
by the fact \cite{chru} that this behaviour is unstable in the class 
of spherically symmetric initial data.

An important new element was brought into the study of the 
Einstein-scalar field and gravitational collapse in general by 
the numerical work of Choptuik \cite{choptuik93}. He took a fixed 
initial datum for the scalar field (which due to spherical symmetry 
determines initial values for the gravitational field) and scaled it 
by an arbitrary constant factor. This gives rise to a family of 
initial data depending on a real parameter $A$. 
For all initial data used in the computations the same picture 
emerged. For values of $A$ corresponding to small data the field 
dispersed, in agreement with the rigorous results. For large 
initial data a black hole was formed.
Under these circumstances we can define a critical parameter 
$A_\ast$ as the lower limit of those values of $A$ for which a 
black hole is formed. If the mass $M$ of the black hole formed is 
plotted as a function of $A$ it is found that $M(A)$ is continuous. 
In particular $\lim_{A\to A_\ast} M(A)=0$ so that black holes of 
arbitrarily small mass can be formed within the given one-parameter
family. Up to now it has not been possible to confirm this behaviour 
by rigorous mathematical arguments. Choptuik's results contain 
much more detailed statements than the one just mentioned, but it 
is the one which will be important in the sequel. 

The nature of the boundary between dispersion and black hole 
formation in gravitational collapse is now a very active research 
area. Most (but not all) of the work has been concerned with 
spherical symmetry and varying matter models and has relied 
essentially on numerical computations. For a recent review of the 
field see \cite{gundlach97}. In a study of critical collapse for 
the Einstein-Yang-Mills equations, Choptuik, Chmaj and Bizo\'n 
\cite{choptuik96} have found both cases where 
$\lim_{A\to A_\ast} M(A)=0$ and 
$\lim_{A\to A_\ast,A >A_\ast} M(A)>0$. They call 
these cases type II and type I respectively, the terminology being
motivated by an analogy to phase transitions in statistical mechanics.
They relate type I behaviour in this system to the existence of the
Bartnik-McKinnon solutions \cite{bartnik88,smoller91}, which are 
static. (The models which had been considered previously, and which 
showed exclusively type II behaviour, admit no regular static 
solutions.)

Almost all the matter models which have been considered up to now 
in the context of critical collapse are field theoretic in nature 
rather than phenomenological. The one exception is a perfect 
fluid with linear equation of state, which shows type II behaviour. 
A phenomenological matter model whose collapse is of interest is 
collisionless matter described by the Vlasov equation. As will be 
described next, the spherical collapse of collisionless matter has 
been studied both analytically and numerically but the known results 
for this type of matter say little about the nature of 
critical collapse. The purpose of this paper is to begin the numerical
investigation of critical collapse of collisionless matter.

In \cite{rein92} it was shown that for sufficiently small initial
data for collisionless matter the matter disperses to infinity at 
large times. Thus the analogue of Christodoulou's small data result 
holds for collisionless matter. The solutions are geodesically 
complete. Unfortunately, no analogue of his large data result 
is known is this case. The only result in that direction which has 
been proved is that there do exist initial data which develop 
singularities \cite{rendall92}. The proof proceeds by demonstrating 
the existence of initial data which contain trapped surfaces and 
applying the Penrose singularity theorem. A different
kind of large data result, which is relevant to the numerical 
calculations of this paper, is that if data given on a hypersurface 
of constant Schwarzschild time gives rise to a solution which 
develops a singularity after a finite amount of Schwarzschild time, 
then the first singularity occurs at the centre of symmetry 
\cite{rein95}.  An analogous result where Schwarzschild time is 
replaced by maximal slicing has also been proved \cite{rendall97}.

It seems plausible that in those solutions which develop singularities
black holes are usually formed. However there are no mathematical 
results on this and the convincing evidence for the formation of 
black holes is purely numerical. This is part of a large body of 
work due to Shapiro, Teukolsky and collaborators which goes far 
beyond the spherically symmetric case. Here only those results 
will be discussed which are directly relevant to this paper. In 
particular only results for spherical symmetry are covered.
In \cite{shapiro85a,shapiro85b,shapiro86} the collapse of various 
configurations of collisionless matter to a black hole has been 
computed numerically. The relative merits for this problem of 
different choices of time coordinate (polar or maximal) and radial 
coordinate (area or isotropic) are discussed in 
\cite{petrich85,petrich86}.

In this paper we carry out an experiment analogous to that of 
Choptuik for spherically symmetric collapse of collisionless matter. 
We use Schwarzschild coordinates (i.e. polar slicing and area radius).
Starting with a suitable smooth function $f_0$ of compact support as 
initial datum for the distribution function we consider the scaled 
data $Af_0$, where $A$ is a positive constant and compute the 
corresponding time evolution numerically. For all data for which we 
tried the experiment, the results were qualitatively
similar. When $A$ is sufficiently small the matter disperses, in 
agreement with the analytic theory. This happens up to some value 
$A_*$ of $A$. For $A > A_*$ the observed behaviour indicates the 
formation of a black hole. The lapse function develops an abrupt 
step at a certain radius $r(A)$.
This step remains at the same radius but gets deeper and deeper. We
interpret this as the signature of a black hole with mass 
$M(A)=r(A)/2$. If $M(A)$ is plotted as a function of $A$ it is 
found that the limiting value of $M(A)$ as $A$ approaches $A_*$ 
from above is strictly positive. Thus we find behaviour of type I 
in the terminology of \cite{choptuik96}. We never find any signs 
of singularity formation for any value of $A$ and
this is consistent with the standard picture where the only 
singularities formed are those of black hole type and they are 
avoided by a Schwarzschild time coordinate.

As a check on the interpretation of the numerical solutions as 
describing collapse to a black hole, radial null geodesics were 
computed. The results agreed well with the expected picture. 
Radial null geodesics starting at the centre at early times escape 
to large values of $r$. Those starting after a certain time $T_1$ 
remain within a finite radius. The limit of this radius as $t$ 
tends to $T_1$ from above is equal to $r(A)$ as computed
above. Thus we obtain a consistent picture with a black hole whose 
event horizon is generated by the null geodesics starting at the 
centre at time $T_1$.

To have a better picture of what is happening in the collapse of near
critical initial data it is useful to consider how the mass $M(A)$ 
of the black hole formed depends on the ADM mass of the initial 
configuration. For parameter values $A$ well above $A_*$ these 
quantities are almost equal. In other words, essentially all the 
matter falls into the black hole. As the critical parameter value 
is approached from above some of the matter does not fall into the 
black hole and in some of the cases which were computed escapes  
to infinity. However, in all the cases which were computed the mass 
of the black hole is more than 90\% of the total ADM mass of the 
configuration. Thus whenever a black hole is formed almost all the 
matter falls into it and the mass gap is a reflection of this. 
The picture in the only other case of a phenomenological matter 
model in which critical collapse has been studied up to now, namely
a perfect fluid with linear equation of state, is very different. 
In the case of a radiation fluid (equation of state $p=(1/3)\rho$) 
the slightly supercritical collapse can be described as follows 
\cite{evans94}. The matter divides almost completely into two parts, 
separated by a near vacuum region. The outer part of the matter, 
which contains almost all the mass, escapes to infinity. The inner 
part, which contains only a very small amount of mass, 
collapses to form a (small) black hole. As the critical parameter is 
approached this situation becomes more and more extreme and the 
black hole mass tends to zero.

The numerical code used is based on a numerical scheme for the 
corresponding Newtonian problem described in \cite{schaeffer87}. 
It incorporates less refined features than the codes of Shapiro 
and Teukolsky but seems to be quite sufficient for the present 
task. It does have the advantage that an analogous Newtonian code 
has been proved to be convergent \cite{schaeffer87} and it seems 
reasonable to hope that this proof could be 
extended so as to obtain results on the convergence of the method 
used in this paper in the future.

These results add collisionless matter to the class of matter models 
for which something is known about critical collapse. Clearly it is 
desirable to examine other types of initial data so as to discover 
the prevalence of the type of behaviour found here or of others 
which have not yet been
seen. It is our hope that these numerical investigations can also 
help to further the mathematical study of collisionless matter in 
general  relativity by providing pictures of what is happening which 
can suggest
which theorems one should try to prove and by what means. Mathematical
investigations of partial differential equations often proceed by
estimating the growth rates of certain quantities and it is useful
to have an idea of the expected growth rates on the basis of 
numerical computations. The fact that we observe no singularities in
the numerical computations can be seen as evidence that the weak
cosmic censorship conjecture is true for collisionless matter. Indeed
it may even be true in a stronger version than in the case of the 
massless scalar field. There may be no naked singularities formed 
for any regular initial data rather than just for generic initial 
data. This speculation
is based on the fact that the naked singularities which occur in 
scalar field collapse appear to be associated with the existence 
of type II critical collapse.

The paper proceeds as follows: In the next section we formulate the
Vlasov-Einstein system, first in general coordinates, and then in
coordinates adapted to the spherically symmetric, asymptotically flat
situation that we want to study. 
In Section 3 we describe the code we are using and test it on a
steady state. In Section 4 we present the results of the
numerical simulations. 

\section{Formulation of the spherically symmetric Vla\-sov-Einstein 
system}
\setcounter{equation}{0}

In the present paper the matter model is a collisionless gas
as described by the Vlasov or Liouville equation. Coupling this 
equation 
self-consistently to the Einstein field equations results in the
Vlasov-Einstein system, which we first write down in general
coordinates on the tangent bundle $TM$ of the spacetime manifold $M$:
\[
p^\alpha \partial_{x^\alpha} f - \Gamma^\alpha_{\beta \gamma}
p^\beta p^\gamma  \partial_{p^\alpha} f = 0,
\]
\[
G^{\alpha \beta} = 8 \pi T^{\alpha \beta},
\]
\[
T^{\alpha \beta}
= \int p^\alpha p^\beta f \,|g|^{1/2} \,\frac{d^4 p}{m} .
\]
Here $f$ is the number density of the particles on phase-space,
$\Gamma^\alpha_{\beta \gamma}$ and $G^{\alpha \beta}$
denote the Christoffel symbols and the Einstein tensor
obtained from the spacetime metric $g_{\alpha \beta}$,
$|g|$ denotes its determinant,
$T^{\alpha \beta}$ is the energy-momentum tensor generated by $f$,
$x^\alpha$ are coordinates on $M$,
$(x^\alpha,p^\beta)$ the corresponding coordinates on the tangent
bundle $TM$, Greek indices run from 0 to 3, and
\[
m=|g_{\alpha \beta} p^\alpha p^\beta |^{1/2}
\]
is the rest mass of a particle at the corresponding phase-space point.
We assume that all particles have rest mass 1 and move 
forward in time so that the distribution function $f$ lives
on the mass shell 
\[
PM = \Bigl\{ g_{\alpha \beta} p^\alpha p^\beta = -1,\ p^0 > 0 \Bigr\}.
\]
We consider this system in the asymptotically flat and spherically 
symmetric case and use Schwarzschild coordinates to coordinatize 
the spacetime manifold. The metric takes the form 
\[
ds^2 =-e^{2\mu (t,r)} dt^2 +e^{2\lambda (t,r)}dr^2+ 
r^2 (d\theta^2 + \sin^2\theta d\phi^2),
\]
where $t \in \R,\ r \geq 0,\ \theta \in [0,\pi],\ \phi \in [0,2\pi]$.
Asymptotic flatness is then expressed as the boundary condition
\[
\lim_{r \to \infty} \lambda (t,r) = \lim_{r \to \infty} \mu (t,r) = 0.
\]
We also require a regular centre, which is guaranteed by the boundary 
condition
\[
\lambda (t,0)=0 .
\]
To write the Vlasov equation we use the corresponding Cartesian 
coordinates 
\[
x=(r\sin\phi\cos\theta,r\sin\phi\sin\theta,r\cos\phi)
\]
as spatial and  
\[
v^a = p^a + (e^\l - 1) \frac{x \cdot p}{r} \frac{x^a}{r},\ a=1,2,3
\]
as momentum coordinates. The Vlasov-Einstein system then takes the
form
\be \label{v}
\partial_t f + e^{\m - \l}\frac{v}{\vo}\cdot \partial_x f -
\left( \dot \l \frac{x\cdot v}{r} + e^{\m - \l} \m'
\vo \right) \frac{x}{r} \cdot \partial_v f =0,
\ee
\bea 
e^{-2\l} (2 r \l' -1) +1 
&=&
8\pi r^2 \r , \label{f1}\\
e^{-2\l} (2 r \m' +1) -1 
&=& 
8\pi r^2 p, \label{f2} 
\eea
\bea
\r(t,r) 
&=& 
\r(t,x) = \int \vo f(t,x,v)\,dv ,\label{r}\\
p(t,r) 
&=& 
p(t,x) = \int \left(\frac{x\cdot v}{r}\right)^2
 f(t,x,v)\frac{dv}{\vo} .  \label{p}
\eea
Here $x,v \in \R^3$, $r = |x|$, $\cdot$ denotes the Euclidean dot 
product in $\R^3$, and $\dot \l = \partial \l/\partial t$ and 
$\m'=\partial\m/\partial r$.
$f$ is assumed to be spherically symmetric in the sense that
\[
f(t,x,v)=f(t,Ax,Av),\ A \in \mathrm{SO}(3).
\]
Eqns. (\ref{f1}) and (\ref{f2}) are the $00$- and $11$-components of 
the field equations; it can be shown that for a solution of the above
system also the remaining nontrivial components of the field 
equations hold. We state the $01$-component explicitly, since it 
is used in our numerical scheme:
\be
\dot \l = 
- 4 \pi r e^{\l + \m} j, \label{f3}
\ee
where
\be
j(t,r) 
=
j(t,x) = \int \frac{x\cdot v}{r} f(t,x,v) dv. \label{j}
\ee
This form of the spherically symmetric Vlasov-Einstein system is
convenient for analytical work and has been used in
\cite{rein92,rein95}. For numerical work one wishes to use the 
symmetry explicitly in the Vlasov equation in order to reduce the
latter's dimension. One set of independent variables to use is
\[
r =|x|,\ u = |v|,\ \alpha = \cos^{-1}\frac{x\cdot v}{r u}.
\]
However, an equivalent and more convenient
set of variables is
\be \label{rwf}
r =|x|,\  w = \frac{x\cdot v}{r},\ L = |x|^2 |v|^2 - (x \cdot v)^2 = 
|x \times v|^2,
\ee 
particularly because $L$ is constant along the characteristics of 
(\ref{v}); note that
\[
u^2 = w^2 + \frac{L}{r^2}.
\]
In these variables the Vlasov equation for $f = f(t,r,w,L)$ becomes
\be  \label{vrwf}
\partial_t f +
e^{\m - \l} \frac{w}{\sqrt{1 + u^2}} \partial_r f   -
\left( \dot \l w + e^{\m - \l} \m' \sqrt{1 + u^2} 
- e^{\m - \l} \frac{F}{r^3 \sqrt{1 + u^2}} \right)
\partial_w f = 0. 
\ee
The field equations remain unaffected,
and the source terms (\ref{r}), (\ref{p}), (\ref{j})
can be rewritten in terms of $(r,u,\alpha)$ or $(r,w,L)$.

We now mention some results that have been established for the 
spherically symmetric Vlasov-Einstein system. Throughout
we consider as initial condition a spherically symmetric,
nonnegative function $\fn$ which as a function of $x$ and $v$ is 
continuously differentiable, has compact support and satisfies the 
inequality
\be \label{nots}
\int_{|x|\leq r}\int \sqrt{1+v^2} \fn (x,v)\,dv\,dx < \frac{r}{2},\ 
r > 0,
\ee
which means that the initial hypersurface does not contain a trapped
surface. In \cite{rein92} it was shown that for such an initial 
condition there exists a unique, continuously differentiable solution
$f$ with $f(0)=\fn$, which exists with respect to Schwarzschild time
on some right maximal interval $[0,T[$. If the solution blows
up in finite time, i.~e., if $T<\infty$ then $\rho(t)$ becomes 
unbounded as $t \to T^-$. Actually, as shown in \cite{rein95}, 
$\rho$ in this case
has to become unbounded at the centre $r=0$, i.~e., if any singularity
evolves, the first one must be at the centre. This rules out 
singularities of shell crossing type, which can be a nuisance
in other matter models, e.~g.\ dust. If the initial datum
is uniformly small the resulting solution is global in the sense that
the spacetime is geodesically complete and the components
of the energy momentum tensor as well as metric quantities decay
with certain algebraic rates in $t$.

Let us denote
\be \label{mf}
m(t,r)=4 \, \pi\, \int_0^r s^2 \rho(t,s)\, ds.
\ee
Then $m(t,\infty)$ is a conserved quantity, the ADM mass of the 
system. Using $m(t,r)$ the field equations (\ref{f1}) and (\ref{f2}) 
yield
\be \label{lambda}
e^{- 2\lambda (t,r)} = 1-\frac{2m(t,r)}{r},
\ee
\be \label{mupr}
\mu'(t,r)=e^{2\lambda (t,r)}\left(\frac{m(t,r)}{r^2} 
+ 4 \pi r p(t,r)\right),
\ee
also
\be \label{lambdapr}
\lambda'(t,r)= \frac{1}{2r}\left(
e^{2\lambda} \left( 8 \pi r^2 \r(t,r) - 1 \right) - 1 \right) 
= e^{2\lambda}\left(- \frac{m(t,r)}{r^2} + 4 \pi r \r(t,r)\right);
\ee
note that the right hand side of (\ref{lambda}) is positive
at $t=0$ by the assumption (\ref{nots}). A further quantity which is
conserved by the system is the total number of particles
\be \label{numpart}
\int\int e^\lambda f(t,x,v)\,dv\,dx .
\ee
\section{Description and Testing of the Code}
\setcounter{equation}{0}
Let us consider an initial condition, $\fn(x,v)$, which is spherically
symmetric, satisfies the condition (\ref{nots}),
and vanishes outside the set $(r,u, \alpha ) \in \left[
R_0,R_1\right] \times \left[ U_0,U_1 \right] \times 
\left[ \alpha_0, \alpha _1\right]$. We will approximate the solution
using a particle method.  For a thorough treatment of particle methods
in the context of plasma physics see \cite{bl}.
To generate
the particles we take integers $N_r,N_u, N_{\alpha}$ and define
\[
\Delta r =  
\frac{R_1 - R_0}{N_r},\ 
\Delta u = \frac{U_1 - U_0}{N_u},\ 
\Delta \alpha = \frac{\alpha_1 - \alpha_0}{N_{\alpha}},
\]
\[
r_i =  
R_0 + \left(i-\frac{1}{2}\right) \Delta r, \ u_j = U_0+ 
\left(j-\frac12\right) \Delta u,\ \alpha _k = \alpha_0 + 
\left(k-\frac{1}{2}\right)\Delta \alpha,
\]
\[
f^0_{ijk} =
\fn\left(r_i, u_j, \alpha_k \right)\; 4 \pi r^2_i
\Delta r\; 2\pi u^2_j \Delta u\, \sin \alpha_k \Delta\alpha,
\]
\[
r^0_{ijk} = 
r_i,\ w^0_{ijk} = u_j\,\cos \alpha_k,\ L_{ijk} =
\left( r_i u_j\,\sin \alpha_k \right)^2.
\]
From these, approximations are made of the quantities $\r,\ p,\ j$, 
and $m$ defined in (\ref{r}), (\ref{p}), (\ref{j}), and
(\ref{mf})
at the grid points  $n \Delta r$. The equation 
\be \label{mupr2}
\mu' = \frac{1}{1-\frac{2m}{r}}\left(\frac{m}{r^2} + 4 \pi r p\right),
\ee
which is obtained from (\ref{lambda}) and (\ref{mupr}),
together with $\mu \rightarrow 0 $ as $r\rightarrow \infty$ is used to
compute $\mu$ on this grid.  Note that for $r$ outside the support, 
$p\equiv 0$ and $m \equiv $ constant and this equation is explicitly
integrable.  $\lambda$ is computed using (\ref{lambda}), and similarly
$\dot \l$ and $\l'$ are computed using (\ref{f3}) and 
(\ref{lambdapr}).
Letting $D$ denote differentiation along a characteristic 
of the Vlasov equation (\ref{vrwf}) we have 
\begin{eqnarray*}
Dr 
& = & 
e^{\mu - \lambda} \frac{w}{\sqrt{1+u^2}},\\
Dw 
& = & 
- w \dot\lambda - e^{\mu-\lambda} \sqrt{1+u^2} \mu' + 
\frac{e^{\mu - \lambda} L}{r^3 \sqrt{1+u^2}}\\
DL 
& = & 
0;
\end{eqnarray*}
recall that 
\[
u^2 = w^2 + \frac{L}{r^2}.
\]
We interpolate $\rho,\ j,\ p,\ \mu,\ \lambda$ etc. to particle
locations and use these equations to define $r^1_{ijk}$ and 
$w^1_{ijk}$
using a simple Euler time stepping method.  Here 
$r^1_{ijk}$ denotes an approximation of the radius of the 
characteristic
at time $\Delta t$ with $r = r_i,\ u=u_j,\ \alpha = \alpha _k$ 
at time $0$.

We also have the equation
\[
\left(\Delta t \right)^{-1} \left(f^1_{ijk} - f^0_{ijk} \right) = -
f^0_{ijk} \left( \dot\lambda + w
\lambda ^{\prime} e^{\mu - \lambda} \left( 1 + u^2 \right)^{-\frac 12}
\right)
\]
which represents the time evolution of a volume element along
a characteristic.  One time step is now complete.

To test the code a steady state solution was generated.  
Following \cite{rein93} we take 
\[
f(x,v) = \phi (E)
\]
where 
\[
E = e^{\mu} \sqrt{1 + |v|^2}
\]
is the particle energy. For simplicity 
\[
\phi(E) = \left\{ 
\begin{array}{lll} 1 &, & E < E_0 \\
0 &, & E \geq E_0 
\end{array}\right.
\]
with $E_0 >0$ was taken.  Then from \cite{rein93}, 
\[
\rho(r) = g_{\phi} \left( \mu (r)\right)
\]
and 
\[
p (r) =
h_{\phi}\left( \mu(r)\right)
\]
where 
\[
g_{\phi} (u) = 4 \pi \int^\infty_1 \phi \left(
\epsilon e^u \right) \epsilon ^2 \sqrt{\epsilon ^2-1}\, d\epsilon
\]
and 
\[
h_{\phi}(u) = \frac{4\pi}{3} \int^\infty_1 \phi
\left(\epsilon e^u\right) \left( \epsilon ^2 -1 \right)^{3/2}
d\epsilon.
\]
Substituting into (\ref{mupr2}), $\m$ must satisfy
\be\label{muss}
\mu ^{\prime}(r) = \frac{1}{1-\frac{8 \pi}{r}\int_0^r s^2 
g_\phi (\m(s))\,ds}
\left(\frac{4 \pi}{r^2}\int_0^r s^2 g_\phi (\m(s))\,ds
+ 4 \pi r h_{\phi} (\m(r))\right)
\ee
and 
$\mu \rightarrow 0$ as  $r \rightarrow \infty$.
This was solved using a shooting method.  Note that for $\mu
\geq \ln E_0$ (\ref{muss}) reduces to 
\[
\mu ^{\prime}(r) = \frac{r^{-2}m(\infty)}{1-2r^{-1} m(\infty)},
\]
which is explicitly integrable.  Thus, it was only necessary
to solve (\ref{muss})
on a bounded domain.  $E_0 = 0.9$ proved to be a convenient
choice.  The resulting steady state has mass $3.36 \ 10^{-2}$ and
support contained in $0 \leq r \leq 0.36$.  The radial component of 
$v$ ranges from $-0.64$ to $0.64$.  The maximal values of $\mu$ and
$\lambda$ are $0.296$ and $0.132$ with $\rho$ a decreasing function 
of $r$.

It was found that taking $N_r = 40,\ N_u= 10$, and $N_{\alpha} = 10$
produced good results.  This resulted in 2550 particles (less than
$40\times 10 \times 10$ since the support of $f$ is not
rectangular).  At time zero the maximal errors in $m$ and $\mu$ 
(maximum over $r$) were $1.63 \ 10^{-4}$ and $2.84\ 10^{-3}$ 
respectively. 
Dividing by the maximal values of $m$ and $\mu$ 
(that is, by $3.36 \ 10^{-2}$ and $2.96\ 10^{-1}$) we find the 
maximal errors in $m$ and
$\mu$ are $0.49\%$ and $0.96\%$ at $t=0$.

The errors in $m$ and $\mu $ were computed at time $t=10$ using
different time steps.  
Percentage errors were computed as described
above with the results:

\begin{center}
\begin{tabular}{|c|c|c|}\hline
& & \\
\ \ $\Delta t$ \ \ & \ \ error in $m$ \ \ & \ \ error in $\mu$\ \  \\
& & \\ \hline
& & \\
$\frac{1}{4000}$& $6.2\%$& $5.9\%$ \\
& & \\ \hline
& & \\
$\frac{1}{8000}$ & $3.4\%$ & $2.9\%$ \\
& & \\ \hline
& & \\
$\frac{1}{16000}$ & $2.1\%$ & $1.3\%$ \\
& & \\ \hline
\end{tabular}
\end{center}

Thus the particle code tracks the steady state reasonably
well, although a rather small time step seems to be needed.  We
attribute this, at least in part, to numerical difficulties in 
tracking the motion of particles near $r = 0$.  Note that for this 
steady state the density is largest at $r=0$.  For other solutions 
with zero density near $r = 0$, the time step was taken larger 
without significant change
in the results.

\section {\bf Results of Simulation}
\setcounter{equation}{0}
In this section we consider initial data
\[
\fn (x,v) = A f_0(x,v)
\]
with $f_0$ fixed and vary $A$.  As a first example we take
\[
f_0(x,v) = \Bigl[ 50,000(2.2-r)(r-2)(10.2-u)(u-10)
(3.1 - \alpha)(\alpha - 2.9)\Bigr]^2
\]
for $2 < r < 2.2,\ 10 < u < 10.2,\ 2.9 < \alpha < 3.1$ and
$f_0 (x,v) = 0$
otherwise.  Thus the mass is initially concentrated between
$r=2$ and $2.2$ and is moving inward rapidly.  In most of the 
following simulations the support of $f_0$ is divided into $40$ 
by $20$ by $20$ cells ($40$ in $r$) resulting in $16000$ particles 
and the time step is $0.005$.  The cases where this is not so will 
be pointed out.
\newpage
In figures 1 and 2, $A$ is $0.69$. 
In figure 1 the enclosed mass 
$m(t,r)$ defined in (\ref{mf})
is plotted at times $t=0,\ 2,\ 4$, and $8$.  

\bigskip
\vbox{\vskip270pt
\includegraphics{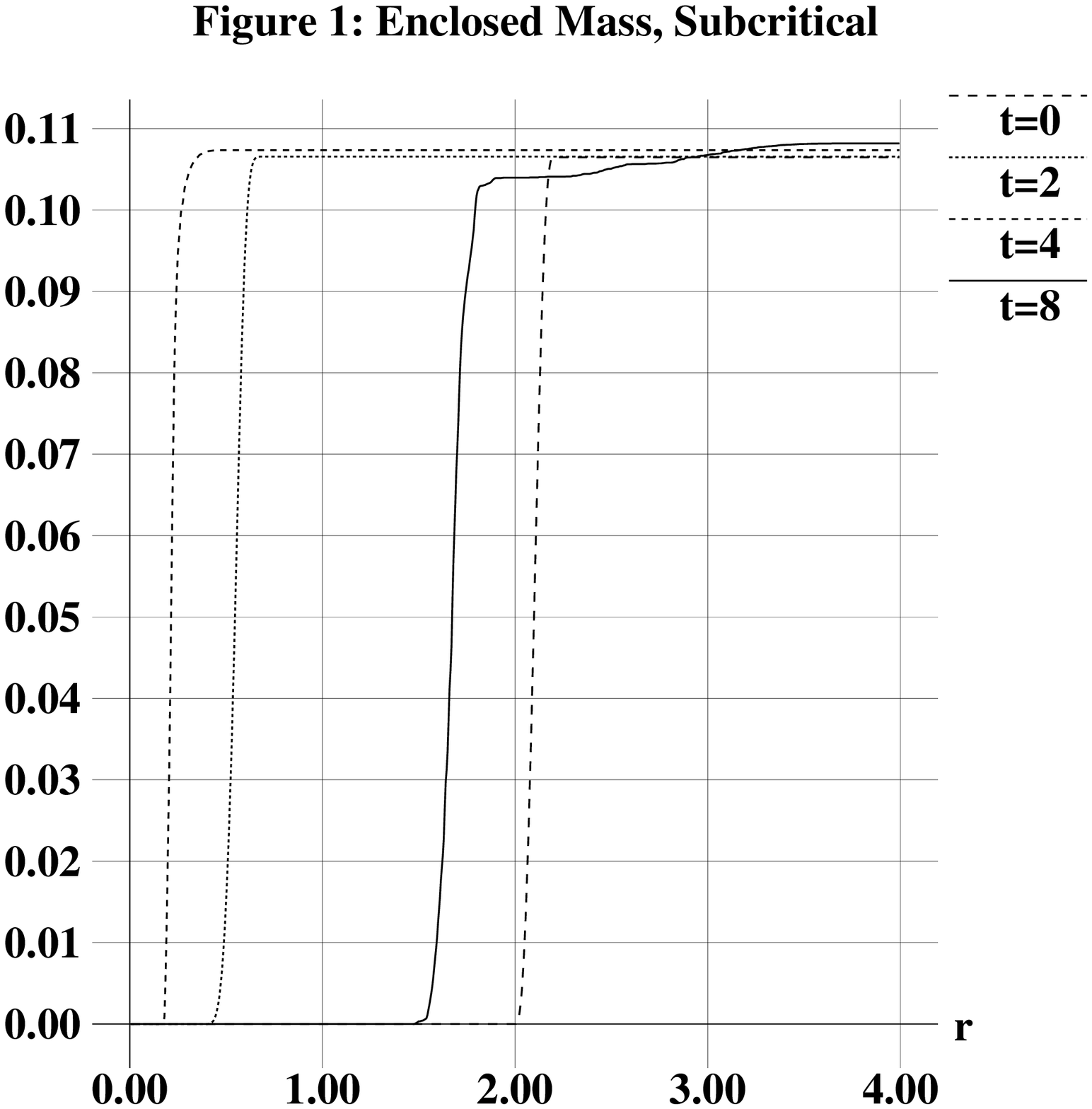}}
\bigskip\noindent
Figure $2$ shows
$\mu$ at the same times.  The solid curve is $t=8$ in both.  
At time $8$
every particle is moving outward with momentum greater than $3.37$ and
has position $r > 1.48$.  In figure 1 we see that the mass has fallen
inward and reversed direction, almost returning to its starting 
position at time $8$.
\newpage  
Comparing figures 1 and 2 we see that at 
$t=4$ the mass is
near $r=0$ and that $|\mu |$ attains relatively large values.  
By $t=8\ \mu$ has dropped to near its starting values.  
As $t$ grows the particles continue outward and disperse, 
consistent with the small data result \cite{rein92}.

\bigskip
\vbox{\vskip270pt
\includegraphics{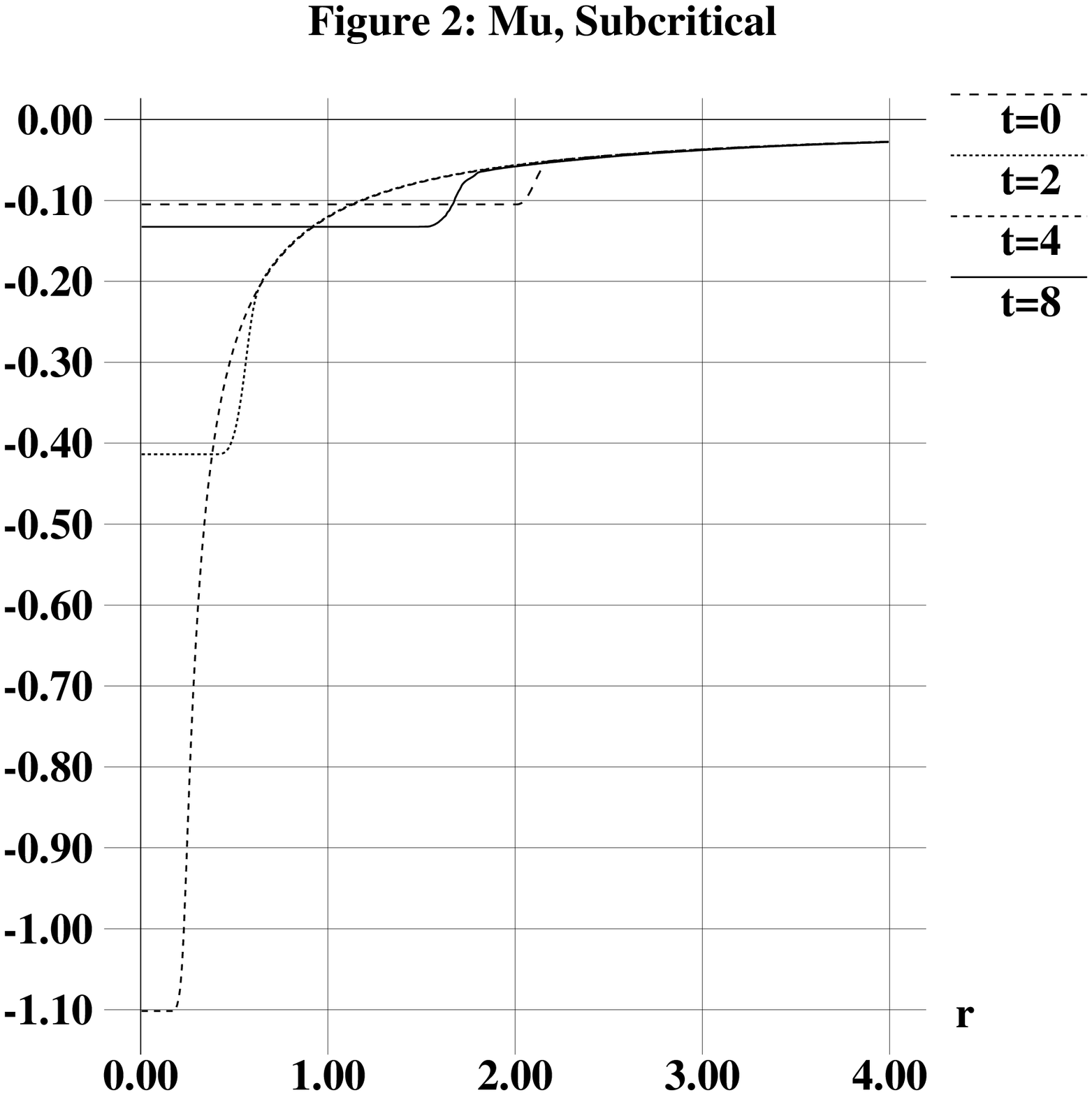}}
\bigskip\noindent
In figures 3 and 4, $A$ is $0.75$.  Again $m$ and $\mu$ are plotted at
times $t=0,\ 2,\ 4$, and $8$ with the solid curve representing $t=8$.
For times $0,\ 2$, and $4$ figure 3 resembles figure 1, but for $t=8$
we see in figure 3 that the mass has not moved back out.  Rather it 
is centered near $0.21$, and from figure 4 $\mu$ has formed an 
abrupt transition near $0.235$.  At time $8$ the maximal outward 
momentum is about $9$ while
the maximal inward momentum is $189$ and the positions satisfy 
$0.13 < r < 0.24$.  Examination of $j(t,r)$ defined in (\ref{j}) 
reveals that the mass flux is almost entirely inward.

\bigskip
\vbox{\vskip270pt
\includegraphics{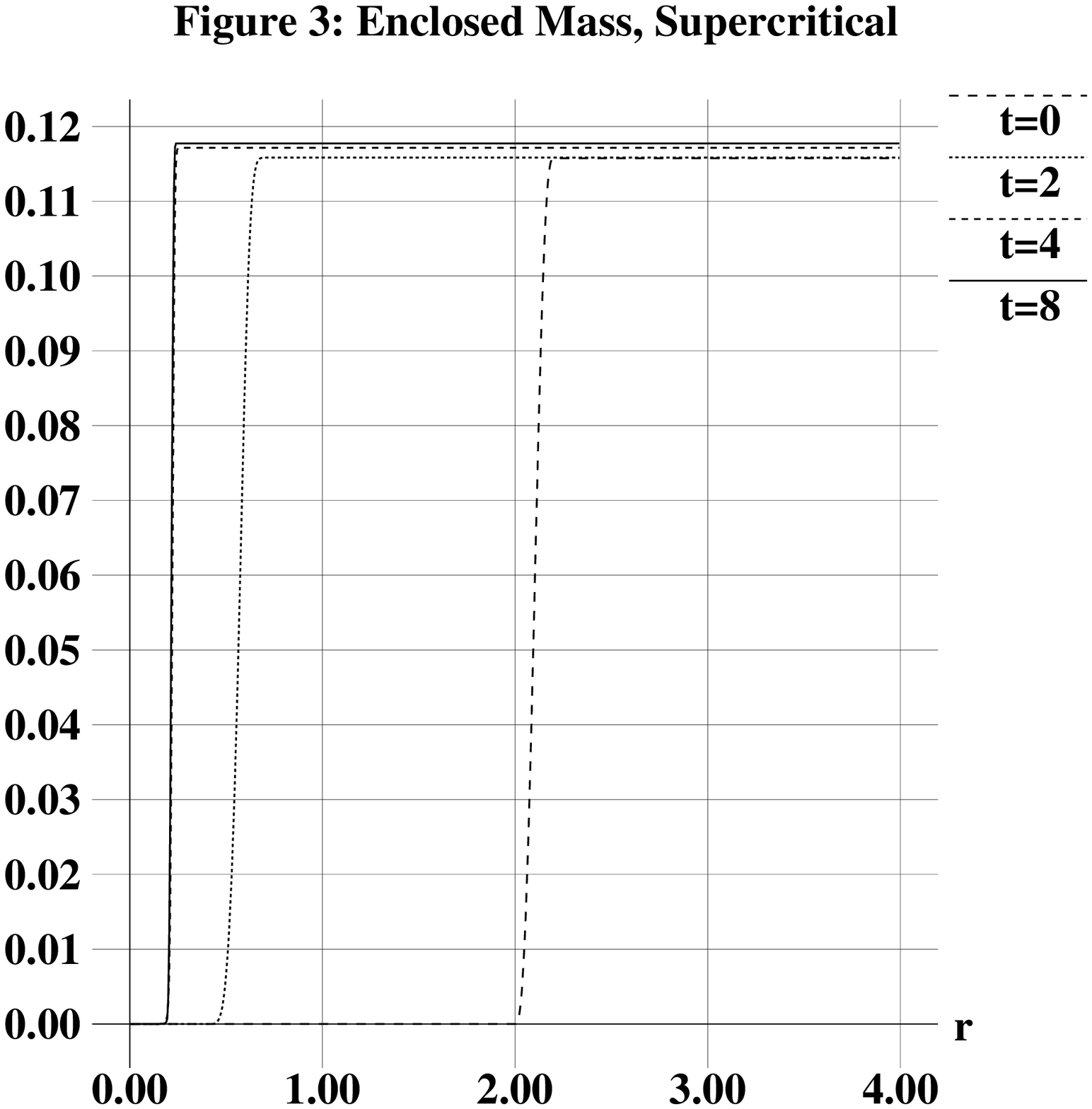}}
\bigskip\noindent
The abrupt transition in $\mu$ occurs at $r= 0.235$ and $\lambda$ has a
maximum at this same $r$.  $\lambda$ forms a cusp at its maximum.  At
time $16$ these features have not moved although the maximal value of
$|\mu |$ has grown.

\bigskip
\vbox{\vskip270pt
\includegraphics{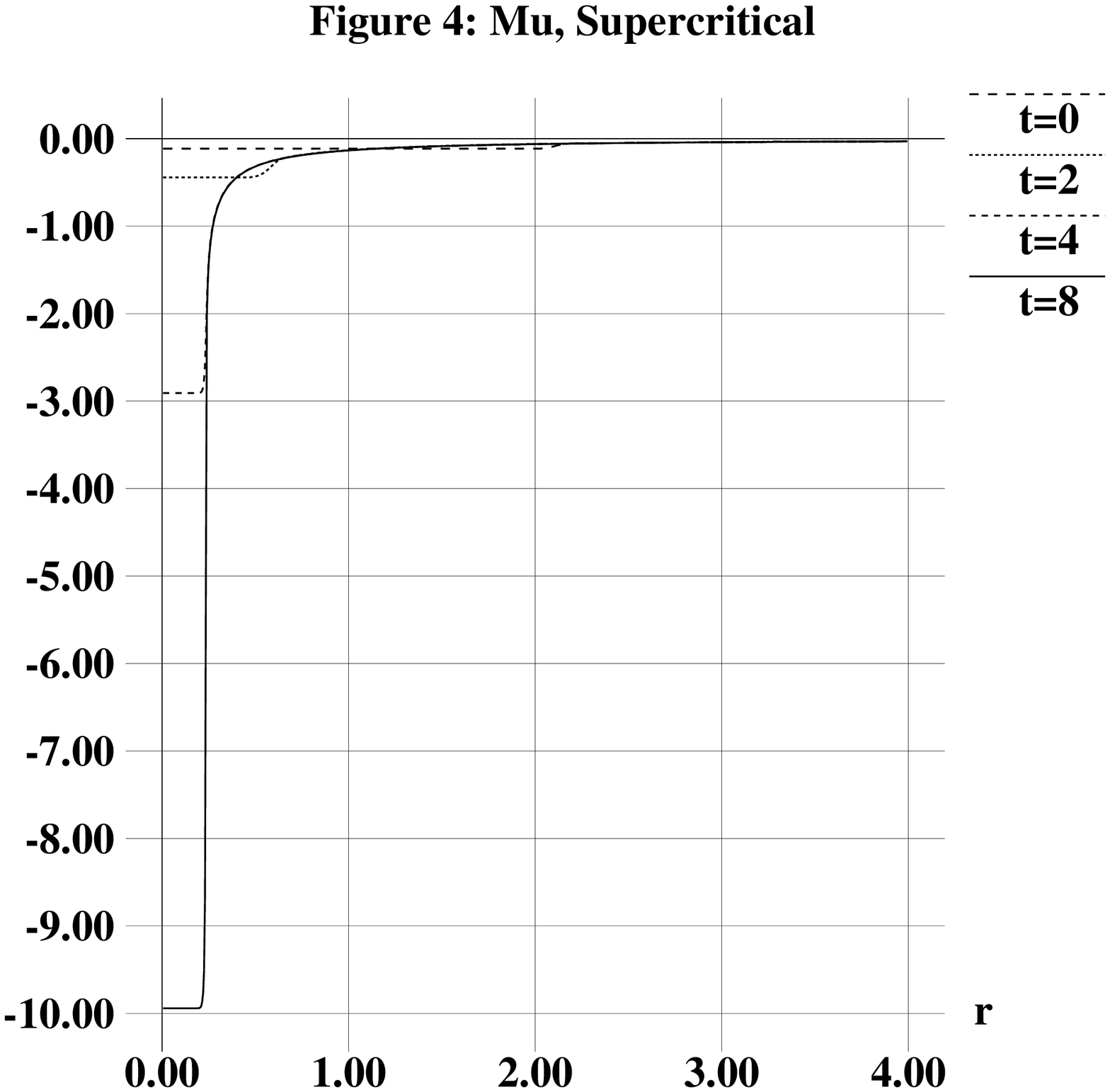}}
\bigskip\noindent
Recall that the quantities 
\[
\int\!\!\int e^{\lambda} f  dv\,dx
\]
and 
\[
\int\!\!\int \sqrt{1+|v|^2} f  dv\,dx
\]
are both conserved by the exact time evolution.  In the above
runs neither quantity varied by more than $1.7\%$ of its initial 
value. A run was made with $80 \times 40 \times 40$ particles, 
$\Delta t = 0.0025,\ A = 0.75$, and final time $8$.  
The resulting graphs of $m$ and
$\mu$ with the finer resolution are qualitatively very similar to
figures 3 and 4, except that the transitions are slightly more abrupt
and the maximal value of $|\mu |$ is increased by about $7\%$.

When $A$ was taken larger than $0.75$ the results are similar.  For
$0.70 \leq A \leq 0.74$ a similar structure formed with a stationary
abrupt transition in $\mu$, but a small amount of mass escaped.  
Thus it seems that the critical value of $A$ for this choice of $f_0$
is $A_\ast \approx 0.70$.

For $A\geq 0.70$ the radius of the maximum of $\lambda$ and the radius
of the abrupt transition in $\mu$ are nearly the same.  Thus we have
computed the radius $r(A)$ where the maximal value (over $r$ and $t$)
of 
\[
\lambda = -\frac{1}{2} \ln \left( 1-\frac{2m}{r}\right)
\]
occurred.  So this is the value of $r$ where the condition 
(\ref{nots}) is most nearly violated.  For $A\geq 0.70$ the 
maximum $\lambda$ was attained at the largest time, for $A\leq 0.69$ 
it was attained earlier. The results are graphed in figure 5.

\bigskip
\vbox{\vskip270pt
\includegraphics{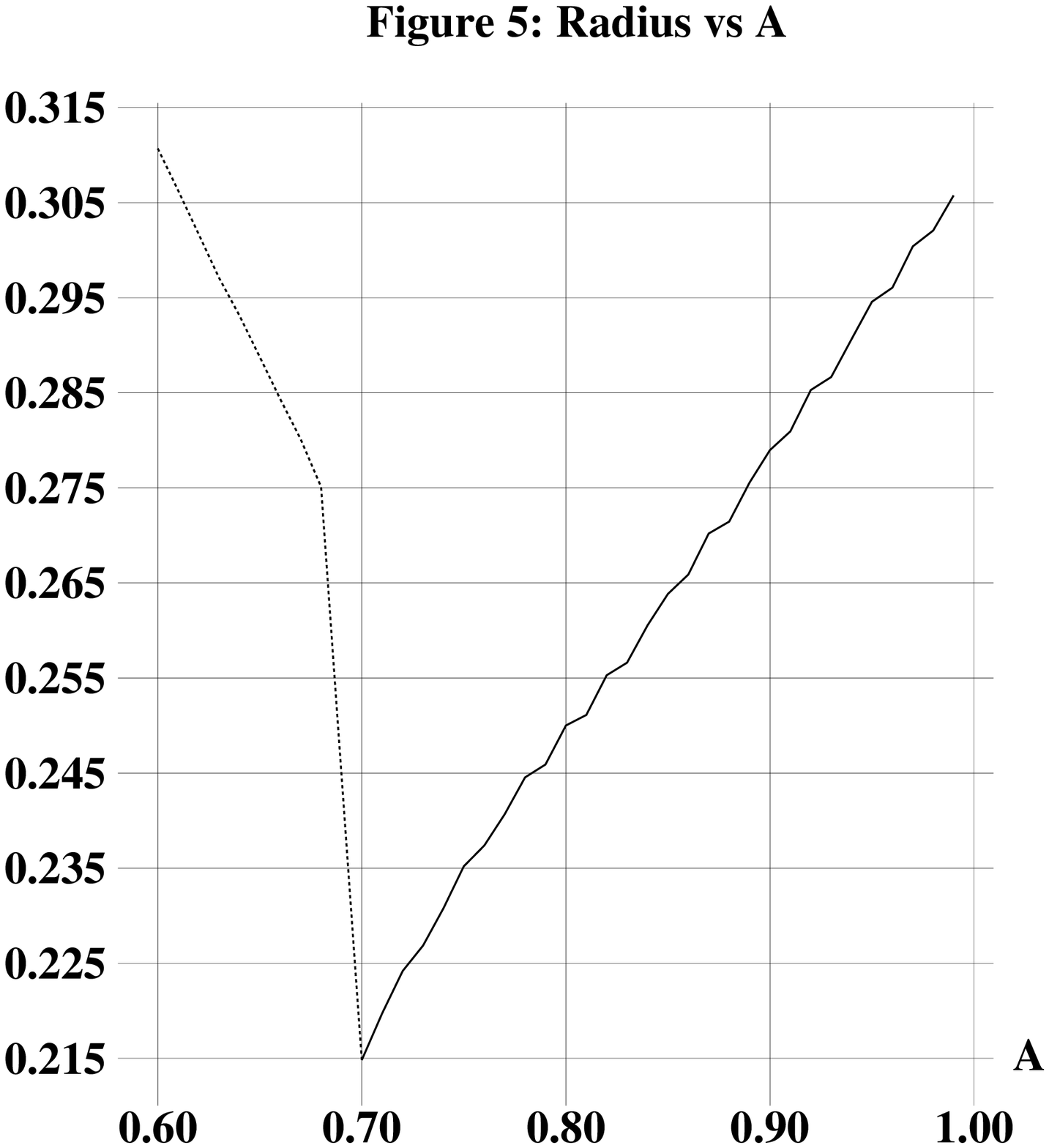}}
\bigskip\noindent
Since the ADM mass of the configuration
depends linearly on $A$ and for $A \geq 0.70$ 
nearly all the mass is captured in the black hole, the mass
of the black hole, $M(A)=r(A)/2$, 
depends nearly linearly on $A$ for $A \geq 0.70$.
The values for $A< 0.70$ are plotted as a dotted curve.
We note that 
\[
\lim_{A\to A_\ast^+} M(A)=\frac{1}{2} \lim_{A\to A_\ast^+} r(A) 
\approx 0.11,
\]
which indicates type I behaviour as explained in the introduction.
Since $\Delta r=0.005$, the discontinuity in $M(A)$ at $A_\ast$ is 
significant.

Next we consider
\[
f_0 (x,v) = 0.1\left(1-r^2\right)^2 \left( 1- u^2 \right)^2
\]
for $r < 1$ and $u < 1$ and $f_0(x,v) = 0$ otherwise.  Similar
to figure 5, figure 6 shows the radius where the maximal value of
$\lambda$ occurs as a function of $A$.  For $A \geq 1.6$ the maximum 
was attained at the largest time.  For this initial condition a 
smaller time step of $0.00125$ was used.

\bigskip
\vbox{\vskip270pt
\includegraphics{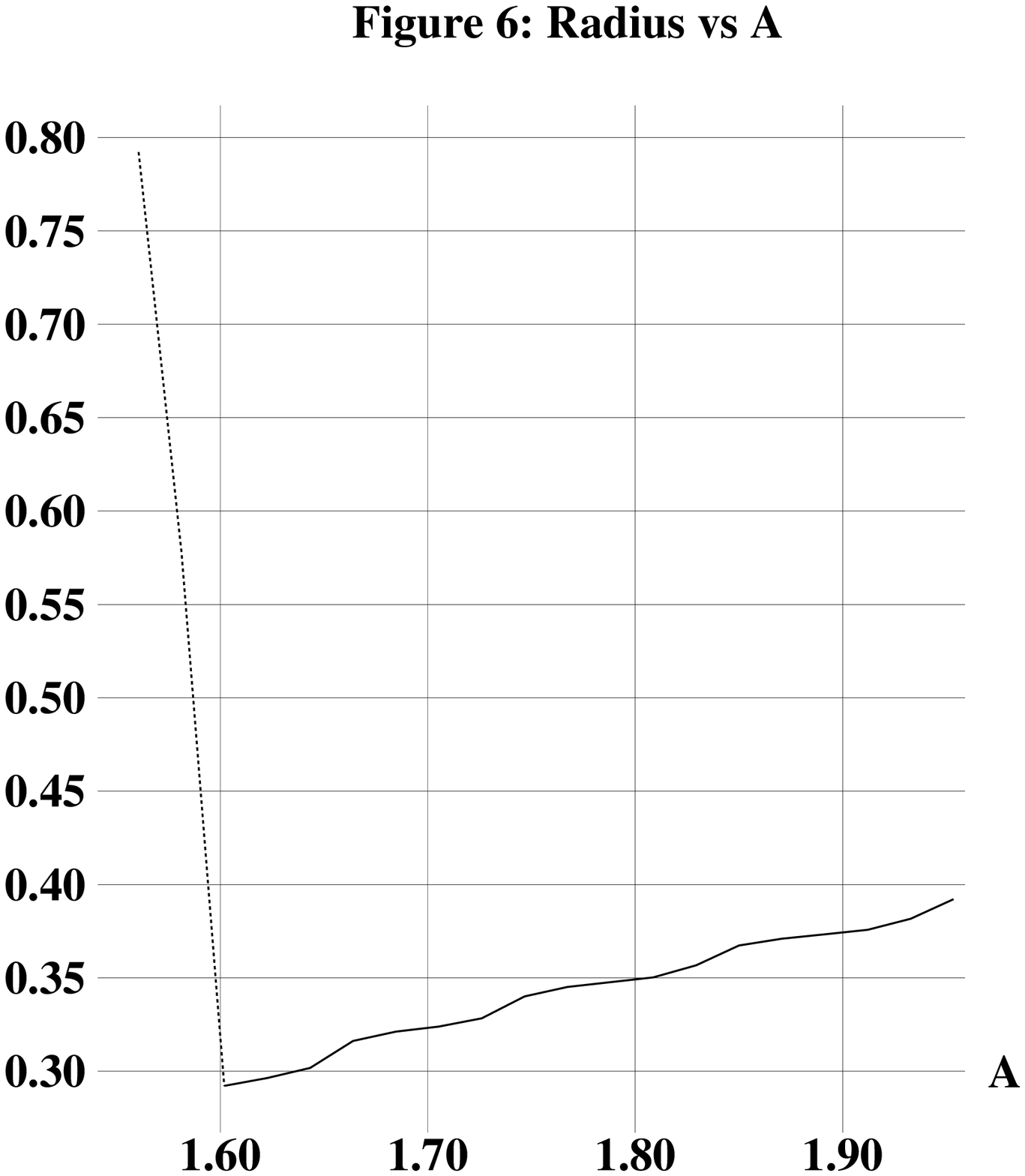}}
\bigskip\noindent
Similarly we consider
\[
f_0(x,v) = 0.1 (3-r)^2 (2-r)^2 (1-r)^2 (1-u^2)^2
\]
for $1 < r < 3$ and $u < 1$ and $f_0(x,v)=0$ otherwise. 
Figure 7 plots radius versus $A$ as figures 5 and 6 did.  
For $A \geq 0.76$ the maximum of $\lambda$ occurred at the 
largest time. For $A \leq 0.75$ the maximum occurred at times near 
zero, hence the nearly constant values of $r$ for $A \leq 0.75$.

\bigskip
\vbox{\vskip270pt
\includegraphics{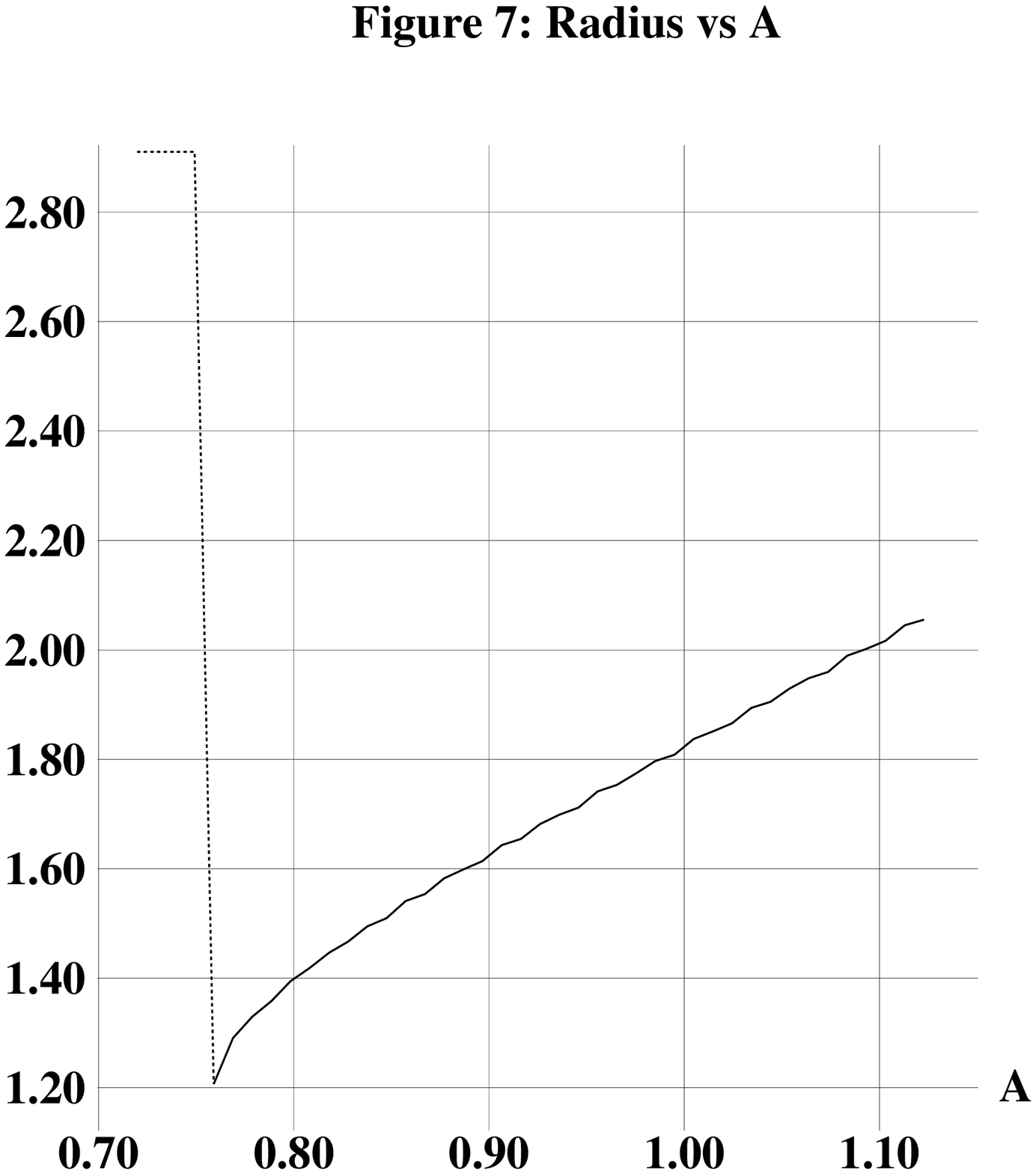}}
\bigskip\noindent
In figures 5, 6 and 7 the final time was taken large enough that
increasing it produced only minor changes.  In each case we see that 
the radius at which $\lambda$
is largest and the step in the lapse function $e^{2\mu}$ forms
remains bounded away from zero.

\noindent
{\bf Acknowledgement:} 
One of the authors (ADR) thanks Carsten Gundlach
for helpful discussions.

\end{document}